\documentclass{article}
\usepackage{caption2}
\usepackage{amssymb}
\usepackage{times}
\usepackage{emulateapj}
\usepackage{psfig}

\def\keV{{\rm\thinspace keV}}

\def\yr{{\rm\thinspace yr}}
\def\kpc{{\rm\thinspace kpc}}
\def\kmps{{\rm\thinspace km}\,{\rm\thinspace s}^{-1}}
\def\arcmin{{\rm\thinspace arcmin}}

\begin{document}

\title{Observations of Abell~4059 with the {\it Chandra X-ray
Observatory}, {\it Hubble Space Telescope} and {\it Very Large
Array}: unraveling a complex cluster/radio-galaxy interaction}

\author{Yun-Young Choi\altaffilmark{1,2,3},
Christopher~S.~Reynolds\altaffilmark{4},
Sebastian Heinz\altaffilmark{5,6,7},
Jessica~L.~Rosenberg\altaffilmark{8},
Eric~S.~Perlman\altaffilmark{9},
Jongmann Yang\altaffilmark{1,2}
}

\affil{$^1$Department of Physics and CSST, Ewha University, Seoul
120-750, Korea}
\affil{$^2$Center for High Energy Physics, Kyungpook National
University, Daegu 702-701, Korea}
\affil{$^3$JILA, Campus Box 440, University of Colorado,
Boulder CO~80309}
\affil{$^4$Dept. of Astronomy, University of Maryland, College Park, MD~20742}
\affil{$^5$Max-Planck-Institut f\"{u}r Astrophysik,
Karl-Schwarzschild-Str. 1, 85740 Garching, Germany}
\affil{$^6$Center for Space Research,Massachusetts Institute of Technology, 77
    Massachusetts Avenue, Cambridge, MA02139}
\affil{$^7$Chandra Fellow.}
\affil{$^8$CASA, Campus Box 389, University of Colorado,
Boulder CO~80309}
\affil{$^9$Dept. of Physics, University of Maryland Baltimore County, 
1000 Hilltop Circle, Baltimore, MD~21250}

\begin{abstract}
  We present a detailed reanalysis of {\it Chandra X-ray Observatory}
  data for the galaxy cluster Abell~4059 and its central radio galaxy,
  PKS2354--35.  We also present new 1.4\,GHz and 4.7\,GHz CnB-array
  radio data from the {\it Very Large Array}\footnote{The VLA is part
  of the National Radio Astronomy Observatory which is a facility of
  the National Science Foundation operated under cooperative agreement
  by Associated Universities, Inc.}, as well as a short archival WFPC2
  image from the {\it Hubble Space Telescope}.  The presence of a
  strong interaction between this radio galaxy and the intracluster
  medium (ICM) was suggested by Huang \& Sarazin (1998) on the basis
  of a short observation by the {\it High Resolution Imager} on {\it
  ROSAT}, and confirmed in our preliminary analysis of the {\it
  Chandra}/ACIS-S data.  In particular, X-ray imaging clearly shows
  two cavities within the ICM that are {\it approximately} aligned
  with the radio-galaxy axis.  However, using our new radio maps
  (which are at lower frequencies and better matched to searching for
  $\sim 1\arcmin$ structures than the previous high-quality maps) we
  fail to find a detailed correspondence between the $\sim 1\arcmin$
  scale radio-lobes and the ICM cavities.  This suggests that the
  cavities are ``ghosts'' of a previous burst of powerful activity by
  PKS~2354-35.  This is supported by detailed, spatially-resolved,
  X-ray spectroscopy which fails to find any shock-heated ICM,
  suggesting that the cavities are evolving subsonically.  We also
  examine the nature of the central asymmetric ridge (or bar) of X-ray
  emission extending for $\sim30\kpc$ south-west (SW) of the cluster
  center that has been noted in these previous analyzes.  We find the
  ridge to be denser and cooler than, but probably in pressure balance
  with, its surroundings.  The thermal evolution of this structure
  seems to be dominated by radiative cooling, possibly enhanced by the
  radio-galaxy ICM interaction.  We discuss several possible models
  for the formation of this SW ridge and find none of them to be
  entirely satisfactory.  In our preferred model, the SW ridge is due
  to radiative cooling induced by an interaction between a
  radio-galaxy driven disturbance and a pre-existing bulk ICM flow.
  The presence of such a bulk flow (with a velocity of $\sim 500\kmps$
  projected on the plane of the sky) is suggested by the off-center
  nature of the pair of X-ray cavities.  Such a bulk flow can be
  created during a cluster/sub-cluster merger --- the presence of a
  prominent dust-lane in the cD galaxy of Abell~4059, ESO~349-G010, is
  circumstantial evidence for just such a merger.
\end{abstract}


\section{Introduction}

There is an increasing realization that the core regions of clusters
of galaxies are complex and dynamic environments.  For some time now,
it has been argued on the basis of data from imaging X-ray telescopes
that the hot intracluster medium (ICM) of the core regions of rich
clusters is radiatively-cooling on timescales shorter than the age of
the cluster.  This gives rise to the phenomenon known as a cooling
flow.  Fabian (1994) gives an extensive review of cooling flows up to
and including constraints from the {\it ROSAT} observatory.  Prior to
the launches of the {\it Chandra X-ray Observatory} and {\it
XMM-Newton}, the X-ray data strongly argued for the inhomogeneous
cooling flow model in which gas in cluster cores cools from X-ray
emitting temperatures down to unobservable temperatures as part of a
multi-phase ICM over a spatially distributed region of the cluster
core.  An obvious mystery, and a strong hint that the real situation
is more complex, was the lack of cool gas (including significant star
formation) observed at other wavebands.

Not surprisingly, the X-ray view of cluster cores became appreciably
more complicated with the launch of {\it Chandra} and {\it
  XMM-Newton}.  With the very high dispersions possible using the
reflection grating spectrometer (RGS) on {\it XMM-Newton}, detailed
emission line spectroscopy of cluster cores became possible for the
first time.  Using these techniques, observations of Abell~1795
(Tamura et al. 2001) and Abell~1835 (Peterson et al. 2001) both
revealed clear evidence for gas cooling from the virial temperature
$kT>4\keV$ down to 1--2\,keV.  In particular, one could isolate and
identify the L-shell emission lines of iron corresponding to gas
spanning this temperature range.  However, very tight upper limits
were set on the amount of gas below 1--2\,keV which were in strong
disagreement with the standard cooling flow model.  In other words,
there is evidence for gas cooling from the virial temperature down to
1--2\,keV, whence it disappears.  This result has been generalized to
a sample of clusters by Peterson et al. (2003).  The explanation for
the temperature floor is still far from clear.  Strong (i.e., order of
magnitude) metallicity inhomogeneities will skew the apparent cooling
function such that gas below 1\,keV cools extremely rapidly, thereby
eluding detection (Fabian et al. 2001).  However, it is not known how
such inhomogeneities will be formed or maintained.  Thermal conduction
and the action of a central radio galaxy may also be important in
producing these temperature floors (Fabian, Voigt \& Morris 2002;
Voigt et al. 2002; Ruszkowski \& Begelman 2002).

In addition to spectral complexity, {\it Chandra} has revealed that
many clusters possess morphological complexities that are thought to
arise due to the interaction of the ICM with a central radio galaxy.
In some cases, the association is clear.  For example, Perseus~A
(Fabian et al. 2000), Hydra~A (McNamara et al. 2000; David et
al. 2001; Nulsen et al. 2002), Abell~2052 (Blanton et al. 2001), and
Cygnus~A (Smith et al. 2001) all show well defined cavities in the
X-ray emitting gas which are coincident with the current radio lobes
of the central radio galaxy.  In these sources, it is clear that the
radio lobes have displaced the X-ray emitting gas producing the
observed X-ray/radio anti-coincidence.  

{\it Chandra} has also revealed the presence of ``ghost'' cavities,
i.e., X-ray cavities that are {\it not} coincident with the active
radio lobes.  Examples include the outer cavity of Perseus~A (Fabian
et al. 2000), Abell~2597 (McNamara et al. 2001), NGC~4636 (Jones et
al. 2002), and Abell~4059 (Heinz et al. 2002).  In these sources, it
is believed that the cavities are associated with old radio lobes
(related to previous cycles of AGN activity).  The low-frequency
(74\,MHz) synchrotron radio emission expected within this scenario
from these old radio lobes has been observed from the ghost cavity of
Perseus-A (Fabian et al. 2002).

Collectively, these observations give rise to several questions.  Most
hydrodynamic models for the formation of these cavities (e.g., Clarke
et al. 1997; Heinz, Reynolds \& Begelman 1998; Reynolds, Heinz \&
Begelman 2001; Reynolds, Heinz \& Begelman 2002, and references
therein) involve the pressure-driven growth of a shock-bounded cocoon.
However, in almost all cases (with NGC~4636 being a notable exception;
Jones et al. 2002), the X-ray shells that bound the observed cavities
are cooler than the ambient ICM, seemingly at odds with the shock
scenario.  It is plausible that the cool shell arises due to the
``lifting'' of lower-entropy material from the cluster core by the
radio galaxy activity (B\"ohringer et al. 1995; Reynolds et al. 2001;
Nulsen et al. 2002), but we would still expect to see some fraction of
the sources in the shock-bounded phase.  More generally, we need to
assess the implications of such data for models of radio galaxy
evolution.  To achieve this goal requires the detailed analysis of
more {\it Chandra} data together with directed numerical simulations.

Abell~4059 was one of the first clusters known to possess X-ray
cavities on the basis of data from the {\it ROSAT} high-resolution
imager (Huang \& Sarazin 1998).  These cavities were approximately
coincident with the radio lobes of the FRI radio galaxy PKS~2354--35,
which is hosted by the cD galaxy at the center of the cluster.  Huang
\& Sarazin (1998) also noted an interesting bar-like feature in the
central regions of the cluster perpendicular to the radio axis.  The
{\it Chandra} Advanced CCD Imaging Camera (ACIS) observation of this
cluster has been previously described by us in Heinz et al. (2002).
In that paper, it was shown that the coincidence between the radio
lobes and the X-ray cavities is not exact, leading to the conclusion
that these are actually ``ghost'' cavities.  It was also suggested
that the complex X-ray morphology (including the central bar) arises
from an interaction of a radio-galaxy driven expanding cocoon and a
pre-existing bulk ICM flow.  Such an ICM flow may result from the
accretion of a galaxy group by the cluster.   

In this paper, we present a detailed reanalysis of the {\it
  Chandra}-ACIS observation of the core regions of Abell~4059.  We
present a spatially-resolved spectral study of the core regions of
this cluster.  We also present new 1.4\,GHz and 4.7\,GHz radio data
from the Very Large Array (VLA) taken with the CnB configuration,
thereby providing a better match providing a better match to the
typical spatial scales characterizing the X-ray cavities.  We confirm
that the arcmin scale radio lobes indeed do not coincide precisely
with the X-ray cavities, especially to the south-east of the center.
We also find that the ridge of emission to the SW of the center is
cooler and denser, but probably in pressure equilibrium, with the
surrounding ICM.  Furthermore, it is determined that the thermal
evolution of this structure must be dominated by radiative cooling.
We discuss various models for the SW ridge, but prefer an explanation
in which it corresponds to shock/compression induced cooling of ICM
caused by interaction of the radio-galaxy driven disturbance with a
bulk ICM flow --- however, such a model may suffer fine tuning
problems.  Finally, we also present an archival {\it Hubble Space
  Telescope} (HST) Wide Field Planetary Camera 2 (WFPC2) image of the
cD galaxy ESO~349--G010.  The presence of a significant dust lane in
this elliptical galaxy suggests that it has accreted a gas rich
companion galaxy within the past $\sim 10^8$\,yrs.  This provides
further circumstantial evidence for the putative cluster/group merger
required to produce the bulk ICM flow.  Section~2 details the {\it
  Chandra}, VLA and HST data reduction, and Section~3 describes our
imaging and spectroscopy investigations.  The observational results
are summarized, and possible models discussed, in Section~4.

Throughout this paper we assume $H_{0}=65$ km s$^{-1}$ Mpc$^{-1}$ and
$q_{0}=0.5$.  Given a redshift of $z=0.049$, this cosmology places
PKS~2354--35 and Abell~4059 at a luminosity distance of 226\,Mpc.

\section{Observations and basic data reduction}

\subsection{X-ray Observations (Chandra)}

A4059 was observed by {\it Chandra} for 24.6 ksec on September 24,
2000 and for 20.1 ksec on January 4, 2001, using the ACIS detector.
The central radio source, PKS 2354$-$35, was centered 1\,arcmin from
the aim point on the S3 back-illuminated ACIS chip (7) such that the
core of the cluster could be imaged entirely by the S3 chip.  Here, we
analyze data from this chip only. The data were filtered so as to
include only events with ASCA grades 0,2,3,4, and 6. The gain for the
first observation was reprocessed using {\tt acis\_process\_events}
with the latest version of the gain file appropriate for data taken
with a focal-plane temperature of $-120^\circ C$ ({\tt
acisD2000-08-12gainN0003.fits}).  Some periods during the first
observation were affected by background flares.  These were removed by
excluding all times for which the background rate exceeded the
quiescent rate by a factor of 1.2.  This left a total exposure time of
10.1\,ksec and 17.7\,ksec for the two observations, respectively.
Using the blank sky background fields of the same part of the detector
we created a background image and spectrum for each observation. Each
background dataset was processed using the same gain file in the A4059
observation and the aspect solutions of each observation were applied
to the background dataset.  All ACIS data reduction was performed
using the {\it Chandra Interactive Analysis of Observations} (CIAO)
version 2.2.1, and spectral analysis was performed using the {\sc
xspec} fitting package.  CIAO 2.2.1 uses Calibration Database (CALDB)
version 2.12.

For both imaging and spectral analyzes, we did {\it not} use the
overall merged event list for the two observations since it does not
contain a valid aspect solution.  This would lead to slight errors in
the corresponding exposure maps and insufficient information to build
correct Response Matrix Files (RMFs) and Ancillary Response Files
(ARFs).  Therefore, for imaging analysis, we also generated separate
flux-calibrated images for each observation and then combined them to
produce a final image.  To produce an image giving the integrated flux
over the full $0.3-8$\,keV band, we first computed flux-corrected
images in several narrow energy bands and then summed those
flux-calibrated images.  This is necessary since the effective area of
the telescope is both energy and position dependent.  To do this, we
built several exposure maps which were weighted according to both
photoelectric absorption and hot diffuse gas emission models for the
incident source spectrum and therefore make it possible to compute the
correct surface brightness having less significant spectral variation
over the entire image. For spectral analysis, we generated separate
source and background spectra (together with the corresponding RMF and
ARF files) for each of the two observations and then analyzed them
simultaneously.  These data reduction steps were facilitated using the
CIAO threads provided by the Chandra Science Center.

Astrometric information was obtained directly from the aspect solution
provided with the {\it Chandra} data release.  There are two
sources in the {\it Chandra} data that can be clearly identified
in other datasets that have accurate astrometry.  Firstly, the
brightest of the small-scale peaks in the centralmost regions of A4059
lie within 0.5\,arcsec of the radio core of PKS2354-35.  Secondly, a
bright X-ray source to the east of the cluster center lies within
1\,arcsec of the 2MASS source 2MASXJ23570418--3448121.  Thus, we
conclude that the absolute {\it Chandra} astrometry for this
observation is correct to within 1\,arcsec.

\subsection{Radio Observations (VLA)}

The pre-existing high-quality radio maps of PKS~2354--35 were taken
with the VLA in its A- and B-array configurations at 8.5\,GHz and
 4.8\,GHz (Taylor et al. 1994).
Such maps may well miss low surface brightness structure on the arcmin
scale, such as could exist within the observed X-ray cavities of
A~4059.  With this motivation, we obtained new VLA data optimized to
study structures on the size scale of the X-ray cavities.  In detail,
we made continuum flux maps of PKS 2354-35 using the VLA-CnB
configuration at 6 and 20 cm. We observed PKS 2354-35 for 37 minutes
at C-band with a 50 MHz bandwidth centered at $\nu = 4.7351$ GHz and
for 31 minutes at L-band with a 50 MHz bandwidth centered at $\nu =
1.4149$ GHz. PKS 2354-35 is a southern source at $\delta$ = $-$34$^h
45^m 30^s$ hence the use of the northern extension, CnB. We were not
able to make polarization maps from these data because the restricted
time range over which the observations were made meant that no
polarization calibrators were available.  These VLA data were reduced
using standard AIPS methods.  The resultant clean beams for natural
weighted maps are fairly elongated: the L-band (1.4\,GHz) beam had a
major axis of $20.3''$ and a minor axis of $14.5''$ while the major
axis of the C-band (4.7\,GHz) beam was $6.2''$ and the minor axis was
$4.0''$.

\subsection{Optical Observations (HST)}

The HST-WFPC2 observed PKS 2354-35 on 2001 April 19 as part of
snapshot proposal 8683 (PI: R. van der Marel).  The observations were
performed in the F814W (broad I) band and the exposure time was 500
seconds, split into two to allow the elimination of cosmic rays.  The
HST data were retrieved from the public archive and combined using the
IRAF/STSDAS task MKDARK, which also allows the elimination of cosmic
rays.  The data were then rotated to a north-up orientation using the
IRAF task IMLINTRAN and registered using WCS information in the header. We
examined both the high-resolution ($0.0455"$/pix) PC image (on which
the host galaxy of PKS 2454-35 lies) as well as the four-chip mosaic,
which was assembled at $0.09965"$/pix using the IRAF/STSDAS task
WMOSAIC.

\section{Image analysis}

\subsection{X-ray morphology}
\label{xray}

\begin{figure*}
\centerline{\hbox{
\psfig{figure=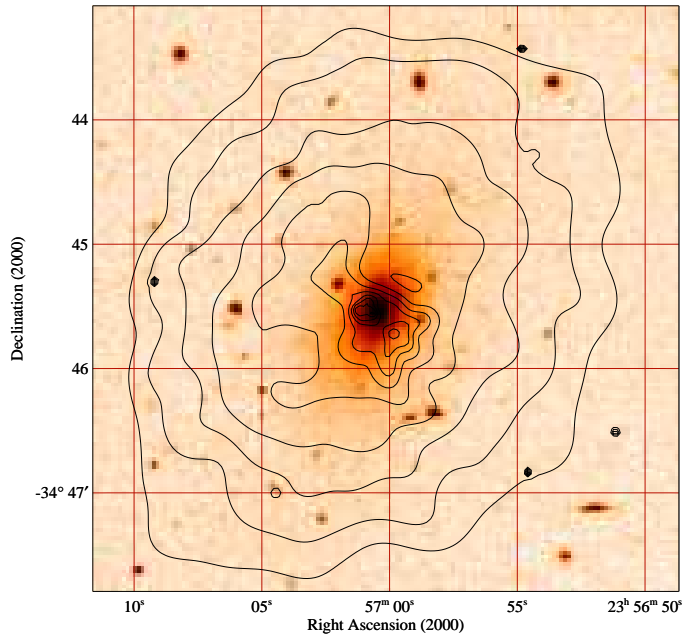,width=0.5\textwidth}
} }
\caption{4-$\sigma$ adaptively smoothed ACIS-S image (contours)
overlaid on the Digitized Sky Survey image of A4059.  See text for
discussion.}
\label{fig:acis_image}
\end{figure*}

For each of the two {\it Chandra} observations, the 0.3--8\,keV data
have been background subtracted and corrected for detector and
vignetting effects using weighted exposure maps (see \S2).  The
resulting processed images were then combined.
Fig.~\ref{fig:acis_image} shows contours of the adaptively smoothed
{\it Chandra} image overlaid on the optical image (from the Digitized
Sky Survey) of the cluster.  The adaptively smoothed image was derived
by smoothing the raw image with a minimum significance of 4-$\sigma$
using the CIAO tool {\tt csmooth}.

It can be seen that the cluster core has a complex X-ray morphology.
The principal morphological features present in these images were
previously noted by Huang \& Sarazin (1998) and Heinz et al. (2002).
The cluster within about $30''$ radius has an hour-glass like
structure (or bar) with two broad peaks.  The strongest peak, at the
center of the cluster, contains further sub-structure with 3 bright
regions. The brightest region coincides with the optical nucleus (ESO
349-G010 from the Digitized Sky Survey), although it is clearly not a
point source.  In fact, we do not detect any pointlike source
coincident with the nucleus of PKS~2354$-$35.  The second of the
principal X-ray peaks is $\sim~15''$ south-west from the center has no
optical or radio counterpart.  The SW edge of this feature is so sharp
as to be unresolved in our adaptively smoothed map; more precisely,
inspection of the smoothing length map produced by {\tt csmooth}
indicates that the SW edge of this feature must be less than
3-4\,arcsec (3-4\,kpc) across.  On larger scales, the X-ray emission
is elongated and aligned along almost the same position angle as the
major axis of the cD galaxy.

Furthermore, there are two cavities in the X-ray emission to the NW
and SE of the cluster center (see Heinz et al. 2002 for a detailed
discussion of the statistical significance of these cavities).  The NW
cavity seen in our Chandra data can clearly be identified with the NW
cavity seen in the ROSAT-HRI data.  For the more subtle SE cavity,
however, the Chandra cavity appears in a different position by about
$20''$ from the ROSAT-HRI cavity.  In order to directly compare the
Chandra image with the ROSAT image, we obtained an image in the
$0.3-2$ keV, which is approximately the band covered by ROSAT.  This
does not change our conclusions regarding the position of the cavity
and the discrepancy with ROSAT.  Considering the unprecedented high
spatial resolution and throughput of Chandra, we conclude that the SE
cavity given by Chandra is likely to be real, not the result of
statistical fluctuation (see Heinz et al. 2002).  After an examination
of the raw ROSAT data, we suggest that a $3\sigma$ fluctuation in the
photon statistics of the ROSAT image may have led to an incorrect
determination of the SE cavity's location.  We also note that it is
difficult to check the correctness of the ROSAT-HRI aspect solution
given the lack of identifiable sources in this short observation.

As pointed out in Heinz et al. (2002), the axis connecting these two
cavities lies perpendicular to the central hour-glass like structure
but the center of the axis does not coincide with the radio galaxy
(see Fig.~\ref{fig:radio} in \S~\ref{radio}).  Prompted mainly by that
fact, Heinz et al. argued that the radio galaxy had interacted with a
moving ICM and, hence, that the cavities had been ``blown'' in the
north-east direction.  As we show in \S~\ref{sec:spat_spec}, spatially
resolved X-ray spectroscopy, as well as the HST-WFPC2 image, provide
further support for this hypothesis.

\subsection{Radio morphology}
\label{radio}

\begin{figure*}
\centerline{\hbox{
\psfig{figure=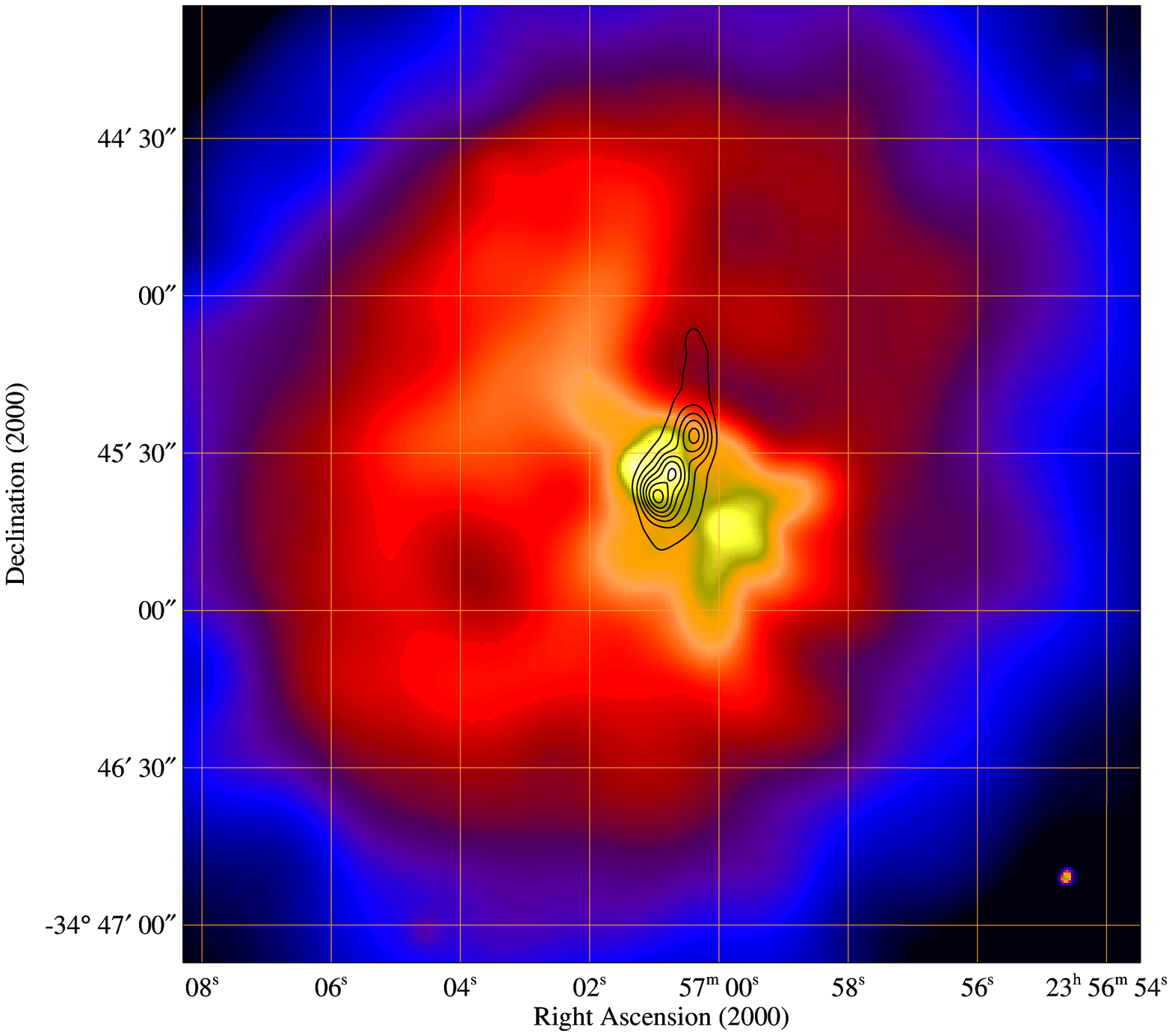,width=0.5\textwidth}
\psfig{figure=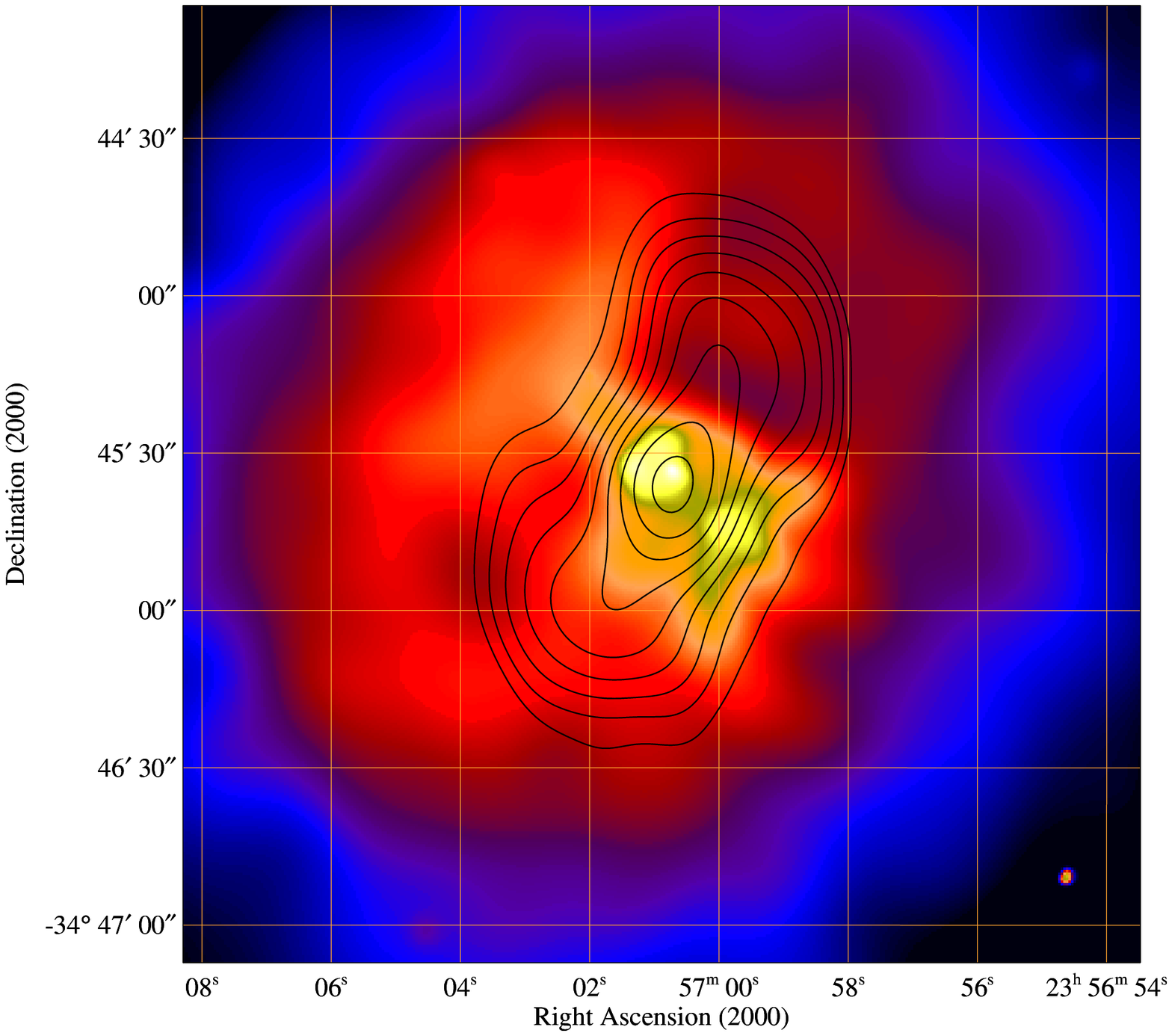,width=0.5\textwidth}
} }
\caption{VLA/CnB-array images (contours) of A4059/PKS~2354-35 taken at
4.7\,GHz (left panel) and 1.4\,GHz (right panel) overlaid on the
4-$\sigma$ adaptively smoothed X-ray image.}
\label{fig:radio}
\end{figure*}

Our new radio maps are presented in Fig.~\ref{fig:radio} as contours
of flux density overlaid on an adaptively smoothed 0.3-8.0\,keV ACIS
image.  It is interesting to compare this with the high-resolution
8.5\,GHz and 4.8\,GHz VLA/A- and B-array images of Taylor et al.
(1994).  While the radio emission at 8.5 GHz extends to the NW X-ray
cavity and coincides with the deepest part of the cavity, it is not
spatially coincide with the SE cavity (see Fig.~1 in Heinz et al.
2002).  Our new CnB-array radio images show that the 1.4\,GHz emission
also does not extend into the SE cavity.  Indeed, the radio axis does
not seem to be aligned with the axis joining two X-ray cavities and
there is no evidence for bending of the radio structure toward the
direction of the two cavities.  Below, we discuss the possibility that
the X-ray cavities correspond to old radio lobes which have been moved
around by bulk ICM flows and may only be directly observable at low
radio frequencies (100\,MHz and below).

\subsection{Optical Morphology}

\begin{figure*}
\centerline{\hbox{
\psfig{figure=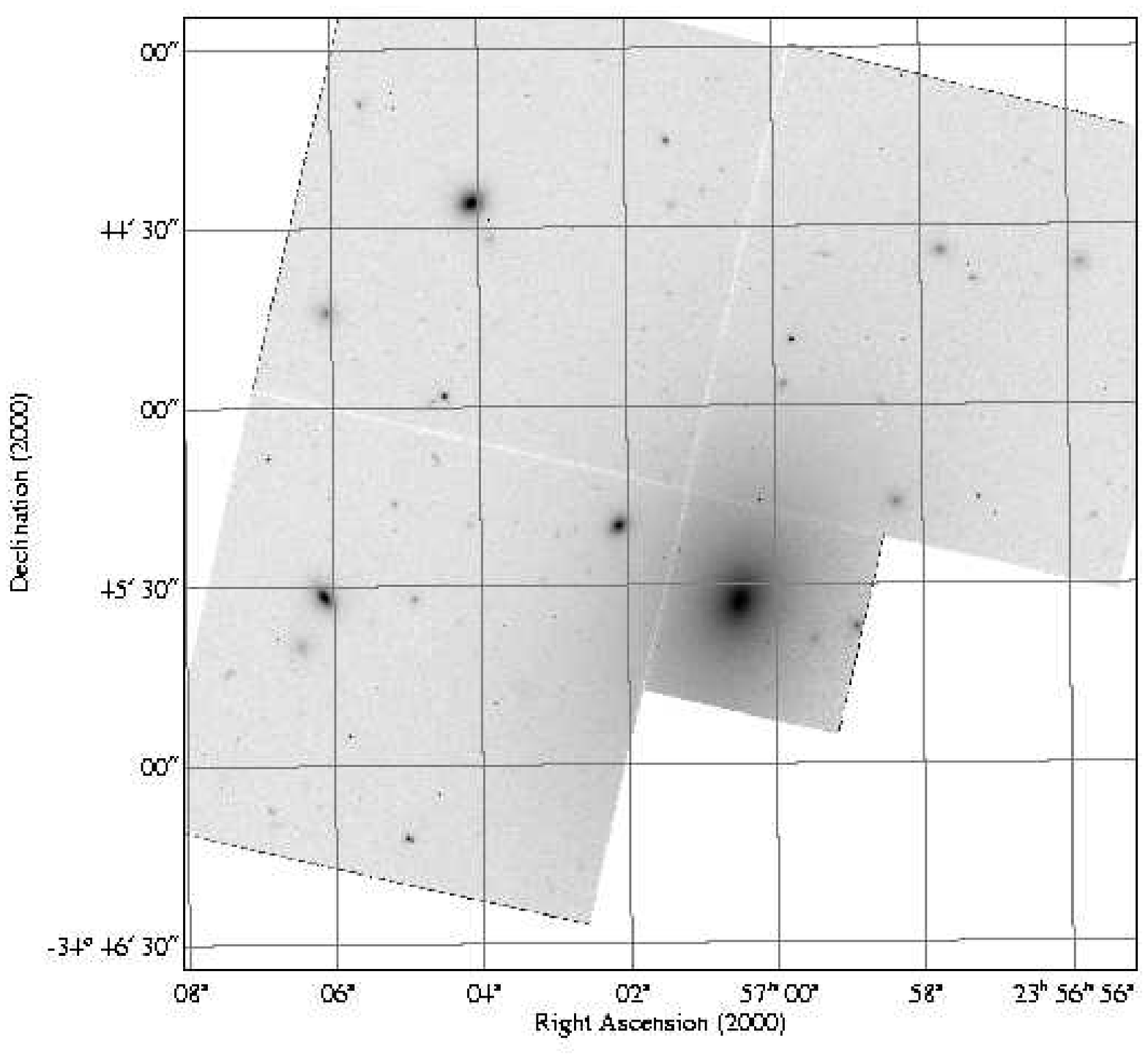,width=0.8\textwidth}
} }
\caption{HST-WFPC-2 image of the A4059 field (see \S2.3 for details of
the observation).  The cD galaxy ESO349--G010 is clearly visible on
the PC-chip.}
\label{fig:wfpc2}
\end{figure*}

\begin{figure*}
\centerline{\hbox{
\psfig{figure=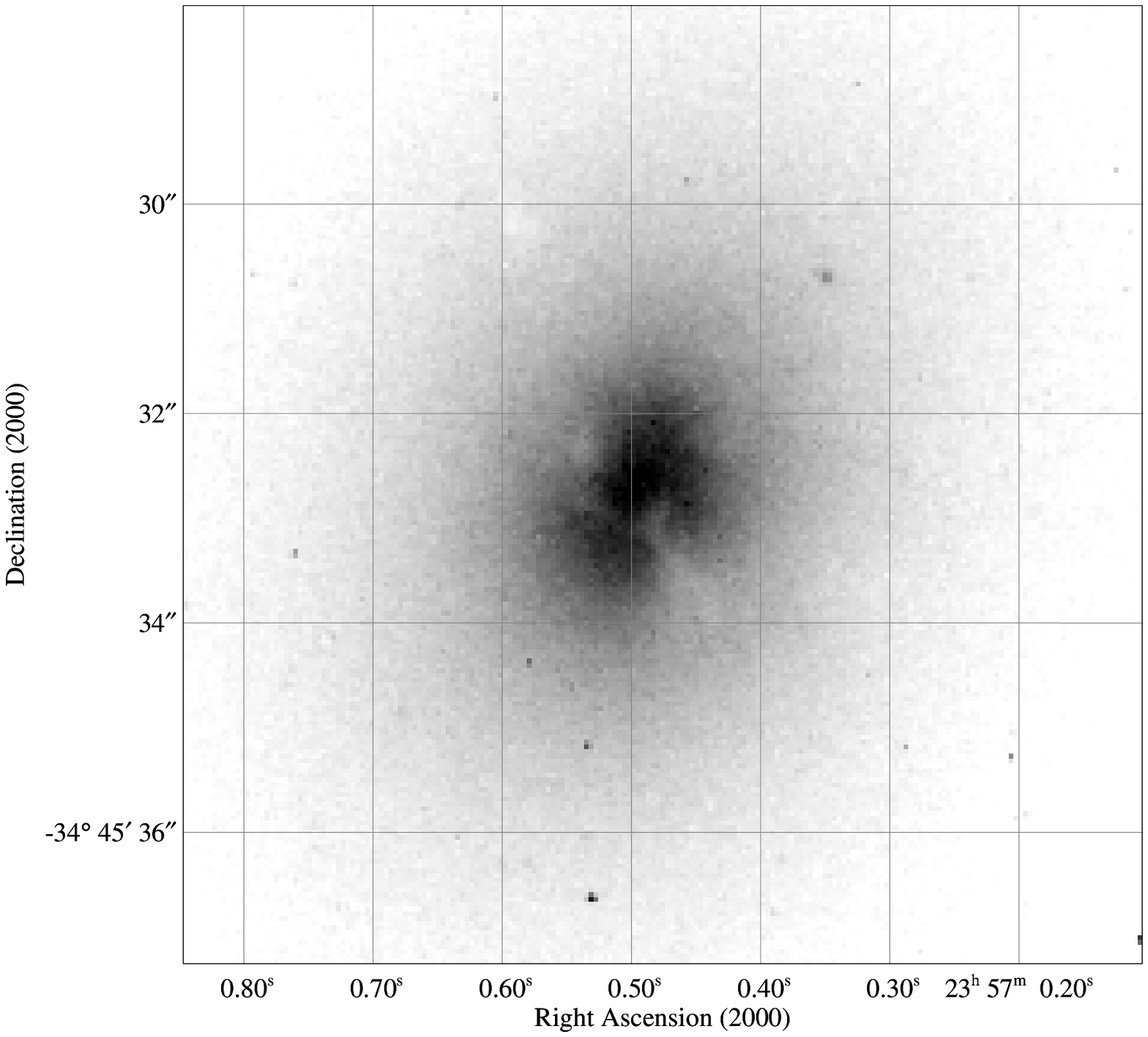,width=0.5\textwidth}
} }
\caption{WFPC-2 imaging of the cD galaxy ESO349--G010, the host galaxy
of the radio source PKS2354--35 (see \S2.3 for details of the
observation).  Note the prominent dust lane crossing the galaxy from
the north-east to the south-west.}
\label{fig:pc}
\end{figure*}

Figure~\ref{fig:wfpc2} shows the HST/WFPC-2 image of the A4059 field.
While an analysis of the optical galaxy cluster is beyond the scope of
this paper, a large number of bright galaxies can be seen.  By far the
brightest, however, is the cD galaxy ESO349--G010 which has been
placed on the PC-chip.  The detailed high resolution PC image of
ESO349--G010 is shown in Fig.~\ref{fig:pc}.  The most striking feature
is the prominent dust lane that can be traced for at least 5\,arcsec
(corresponding to 5\,kpc) projected across the central regions of the
galaxy.  The position angle of the dust lane is twisted by
$60-70^\circ$ relative to the radio axis of PKS2354-35, and coincides
with the position angle of the bright SW ridge in the X-ray image.
Similar dust lanes, also oriented roughly perpendicular to the radio
axis, have been noted in HST data for the host galaxies of three
Compact Symmetric Objects (CSOs) by Perlman et al. (2001).  The
Perlman et al.  CSO-hosts also possess disturbed outer isophotes that
suggest a significant merger event approximately $\sim 10^8\yr$ ago.
Since it was taken as part of a snap-shot campaign, the data for
ESO349--G010 lack the multi-band coverage and signal-to-noise to
perform the Perlman et al. analysis.

\subsection{Spatially-resolved X-ray spectroscopy}
\label{sec:spat_spec}

The X-ray morphology of the core of Abell~4059 is complex and
asymmetric.  Given the obvious lack of spherical symmetry (especially
in the central regions of this system), any spatially-resolved
spectral studies of this cluster will be subject to uncertain
projection effects.  With these caveats in mind, this section presents
a detailed spatially-resolved spectroscopic investigation of
Abell~4059.

\subsubsection{Adaptively binned image analysis}
\label{sec:adap}

\begin{figure*}
\caption{Maps of column density (panel a), temperature (panel b), gas
density (panel c), metallicity (panel d), pressure (panel e), specific
entropy (panel f), cooling time (panel g), and the appropriated
reduced $\chi^2$ value (panel h) for adaptively binned central $2.5'$
image of A4059, with an intrinsic absorption single temperature MEKAL
model.  Dotted circle contour ($r< 25$ \,kpc) includes the bright
hour-glass like structured region of the cluster.}
\label{fig:maps}
\end{figure*}

Given the asymmetries present in the cluster core, we must analyze the
spectral properties of the cluster across the 2-dimensional image.  We
achieve this using the ``adaptive binning'' method of Sanders \&
Fabian (2001).  The adaptive binning code, kindly provided by Jeremy
Sanders\footnote{http://www-xray.ast.cam.ac.uk/~jss/adbin/}, computes
the optimal tiling across the image such that each tile possesses at
least a specified number of photons.  Spectra can then be extracted
and analyzed for each tile.  The major advantage of this method is
that one can maintain high spatial resolution (i.e. small tiles) in
the high count rate regions of the image.

The adaptive binning was set such that each tile possessed at least
600 counts, resulting in a fractional error on the net count rate of
0.04.  Spectra and response matrices were extracted from each tile.
Although an analysis using the $0.3 - 8.0$ keV energy range could
provide information on cold emission and intrinsic absorption in the
cluster, such an analysis is severely hampered by calibration problems
at the lowest energies, in particular the effects of contamination on
the ACIS filter and charge-transfer inefficiency (CTI) in the
CCD. Therefore, only data between $0.5 - 8.0$ \,keV were included in
this spectral analysis.  The separate responses for each tile were
weighted to account appropriately for instrumental response variations
across the detector, using the {\tt mkwarf} and {\tt mkrmf} scripts
implemented within CIAO.  The original auxiliary response files
created by CIAO tool {\tt mkwarf} were corrected for degradation in
the ACIS quantum efficiency (QE) using the software released by George
Chartas and Konstantin Getman
\footnote{http://www.astro.psu.edu/users/chartas/xcontdir/xcont.html}.
Background spectra were generated using the blank sky fields
(Markevitch 2000) for the same part of the detector.  All spectra were
grouped to have at least 20 photons per energy bin, thereby
facilitating the use of $\chi^2$ fitting.  

For our canonical spectral fits, each spectrum was modelled with a
single temperature optically-thin thermal plasma component (modeled
using the MEKAL model as implemented in XSPEC; Mewe, Gronenschild \&
van~den~Oord 1985; Mewe, Lemen \& van~den~Oord 1986; Kaastra 1992;
Liedahl, Osterheld \& Goldstein 1995) with a metallicity fixed at
$Z=0.4$ and absorbed by the Galactic column density of
$N_{H}=1.45\times10^{20}$ cm$^{-2}$.  Once we obtained the best- fit
plasma emission measure and temperature for each bin, we derived the
density, pressure and cooling time assuming that the plasma is
single-phase and has a line-of-sight path length equal to the radial
distance between the center of the bin and the center of the cluster
(following the method of Fabian et al. 2001).  We also studied the
effect of relaxing the metallicity constraint and including the
possibility of intrinsic absorption.

In Fig.~\ref{fig:maps}, we show maps of the best fitting values of
intrinsic absorption (for those fits that relax the absorption
constraint), temperature, gas density, metallicity (for those fits
that relax the metallicity constraint), pressure, entropy
($s=p/n^{5/3}$), radiative cooling time and $\chi^2_\nu$.  We have
overlaid the X-ray contour map to facilitate comparison.  Significant
complexity can be seen in these maps.  Within the centralmost regions
of the cluster (about $30''$ radius, see dotted circle contour in
Fig.~\ref{fig:maps}a), the gas density and pressure dramatically
increase, reaching peak values of $n_{e}\sim 0.11$cm$^{-3}$ and $p\sim
2.7\times10^{-10}$ erg cm$^{-3}$, and the temperature decreases
reaching down to a value of $kT \sim 1.4$\keV.

\begin{figure*}[t]
\centerline{\hbox{
\psfig{figure=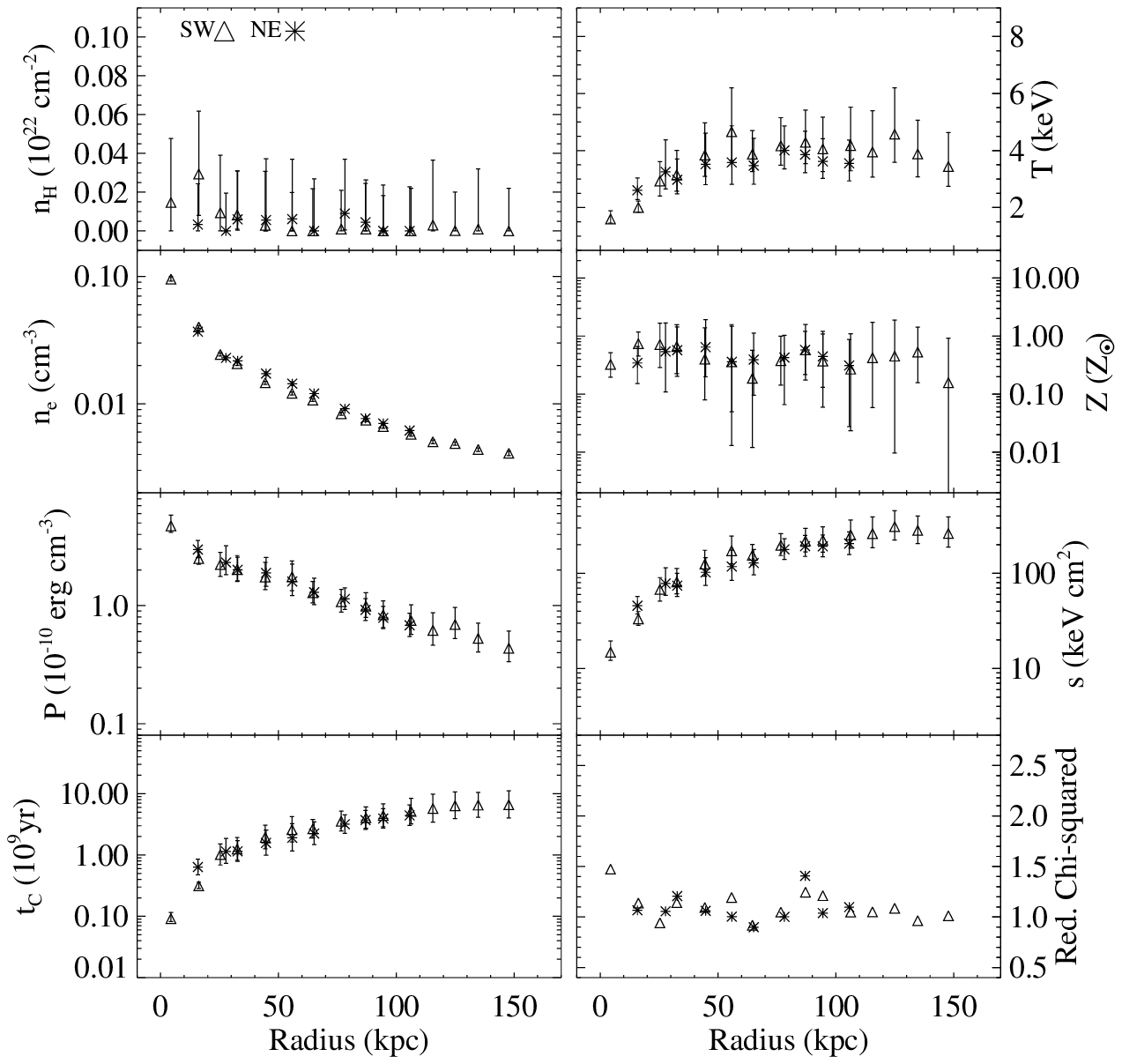,width=0.8\textwidth}
} }
\caption{Radial distribution of the fitted parameters for the bins in
  Fig.~\ref{fig:maps} with 1-$\sigma$ errors.  Each profile
  corresponds to the fitted values of NE and SW sides of the cluster
  center.  The radius is the mean distance from the cluster center to
  the each bin.}
\label{fig:radial}
\end{figure*}

One of the most striking and unusual features within the core of
Abell~4059 is the the bright ridge of emission stretching from the
cluster core to the SW.  Our temperature map (Fig.~\ref{fig:maps}b)
clearly shows that the ridge is composed of gas that is cool (with a
temperature of $\sim 1.4\keV$) and has low entropy.  In order to
investigate the spatial differences in the properties of X-ray
emitting gas in and around this structure, Fig.~\ref{fig:radial} shows
the best fitting parameters for the tile fits, averaged in radial
bins, as a function of the distance from the center of the cluster.
Fig.~\ref{fig:radial} distinguishes between the NE and SW sides of the
cluster in order to study the nature of the bright SW ridge.  The most
significant result from Fig.~\ref{fig:radial} is that the radiative
cooling time within the ridge is rather small (less than 1\,Gyr within
25\,kpc and about 0.1G\,yr within the innermost few kpc).

The temperature and pressure maps in Fig.~\ref{fig:maps}b and
\ref{fig:maps}e exhibit no evidence for any hot gas in or around the
cavities.  We can see that the SW part adjacent to the central
hour-glass like structure shows obviously sharp gradients in the
fitted temperature, entropy, and radiative cooling time maps, while
the NE shows a rather smooth profile.  The oscillation of fitted
values shown in Fig.~\ref{fig:radial} results from this
non-axisymmetric feature.

\subsubsection{Detailed X-ray spectral study of annuli}
\label{sec:proj}

Motivated by the above analysis, we have examined the spectra for
annular regions in the cluster.  Source spectra, background spectra,
response matrices and ancillary files were generated for each annular
region, as in \S~\ref{sec:adap}.  Finally, the spectra were binned so
as to possess a minimum of 20 counts per bin, thereby allowing the use
of $\chi^2$ fitting techniques.  We fitted each spectrum to a variety
of models in the energy range of 0.8--8.0\,keV\footnote{Note that a
  more stringent lower-limit on the energy band (0.8\,keV) is used in
  this analysis as compared with the adaptive-binning analysis
  presented above.  This is to secure against calibration issues in
  these higher signal-to-noise spectra.}: a single-phase emission
model and two multi-phase emission models.  For a single-phase
emission model (hereafter, model-S), the spectrum is fitted with
single-temperature MEKAL model, and for multiphase emission models,
with two-temperature MEKAL model (model-T) or a single temperature
plasma plus cooling flow model (model-SCF).  In the cooling flow
model, we set the upper (initial) temperature of the cooling material
to be equal to the temperature of the single plasma component.  The
lower ``cutoff'' temperature of the cooling flow model is set to
0.1\,keV (i.e., significantly below our bandpass). In this analysis,
the intervening neutral absorption column density is left as a free
parameter.  For reference, the Galactic absorption column density is
$N_{\rm H,Gal} = 1.45 \times 10^{20}$ cm$^{-2}$.

\begin{figure*}[t]
\centerline{\hbox{
\psfig{figure=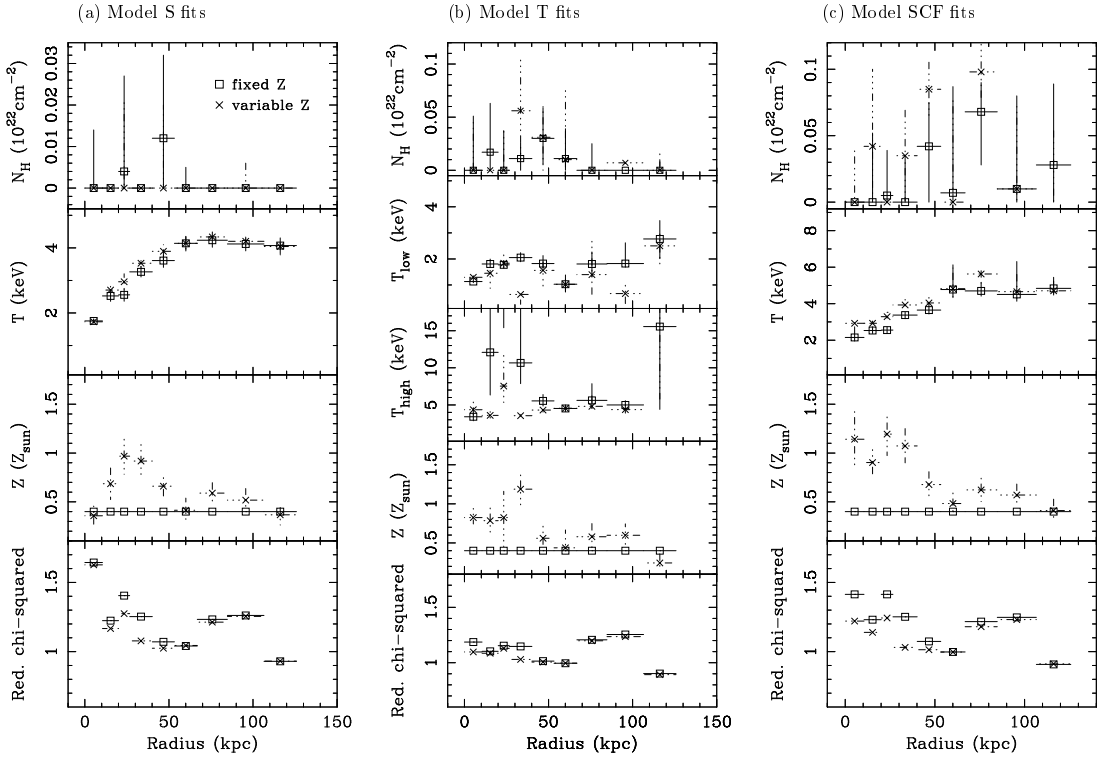,width=0.8\textwidth}
} }
\caption{Temperature, absorbing column density, metallicity, and the
appropriate reduced $\chi^2$ value as a function of radius fitted with
the single temperature MEKAL model (model-S; panel a), two temperature
MEKAL model (model-T; panel b), and single temperature plasma plus
cooling flow model (model-SCF; panel c).  Error bars are shown at the
1-$\sigma$ level for one interesting parameter ($\Delta\chi^2=1$).
Open square and cross show the results with metallicity fixed to the
0.4 times solar value and with metallicity free, respectively.  For
model-T, the metallicities of the two phases were fixed to be same.}
\label{fig:spec}
\end{figure*}

The results of this analysis are shown in Fig.~\ref{fig:spec},
together with the 90\% confidence ranges for one interesting degree of
freedom ($\Delta\chi^2=2.7$).  Here, we report the results for both
the fixed abundance fits (open squares) and variable abundance fits
(diagonal crosses).

When fitted with a single temperature component plasma model
(parameterized by a single temperature and a single emission measure;
Fig.~\ref{fig:spec}a), some clear trends are seen.  The temperature
decreases from 4\,keV in the outer regions of the cluster to 2\,keV in
the central regions.  When applying model-S with metallicity as a free
parameter, we find an enhanced metallicity (approaching almost cosmic
abundances) at intermediate radii (20--50\,kpc), with metallicities
decreasing to $0.4-0.5Z_\odot$ of the cosmic value at smaller and
large radii. The goodness-of-fit is, however, rather poor when
applying model-S to the cluster center.  Much of the poor goodness of
fit is due to an underprediction of the soft flux by the single
temperature model.  It is this mis-match that is responsible for the
unphysically small absorption (i.e. less that $N_{\rm H,Gal}$) implied
by these fits.

The two temperature model (model-T; Fig.~\ref{fig:spec}b) is a much
better description of the spectral data, especially within the inner
50\,kpc.  The actual values of the two temperatures seem to be weak
functions of radius, with $kT_{\rm high}\approx 4-5\keV$ and $kT_{\rm
  low}\approx 1-2\keV$ in most of the radial bins.  Apparent
exceptions to this are the 20--40\,kpc radial bins which, in the fixed
abundance fits, both appear to have $kT_{\rm high}>7\keV$.  However,
variable abundance fits suggest that the abundance strongly deviates
from $Z=0.4Z_\odot$ at these radii and, once that is accounted for,
the upper plasma temperature is also approximately 4\,keV.  The
principal qualitative difference between the one and two temperature
fits lies in the abundance profile.  In the one temperature fits,
there is a pronounced drop in the metallicity as one proceeds from
30\,kpc into the center of the cluster.  On the other hand, the two
temperature fits show a jump in the metallicity at about 40\,kpc, with
the metallicity displaying an approximately flat radial dependence
within this radius.  Thus, the metallicity peak noted in the single
temperature fits is probably an artifact of the model (also see case
of the Virgo Cluster, Molendi \& Gastaldello, 2001).  Due to the
better quality of these fit (especially in the soft band), the
measured absorption column is more meaningful for model-T.  We see
that all radii are consistent with Galactic absorption, i.e., there is
no evidence for intrinsic absorption in this cluster.

The cooling flow model (model-SCF) is a poorer description of these
data than the two temperature model (model-T).  This is due to the
fact that the model includes gas at all temperature from the ambient
temperature down to 0.1\,keV whereas, as noted in the introduction,
the cooling in many clusters (including A4059; Peterson et al. 2003)
is truncated at 1--2\,keV by some process.  With this caveat, we note
that the cooling flow model reproduces the temperature structure of
model-S and the metallicity behaviour of model-T.

\subsubsection{Deprojection analysis}

Of course, the analysis presented in the previous section has not
attempted to correct for projection effects; the observed emission
from a particular annulus contains all of the projected foreground and
background emission, thereby complicating the interpretation of these
results.  

\begin{figure*}[t]
\centerline{
\psfig{figure=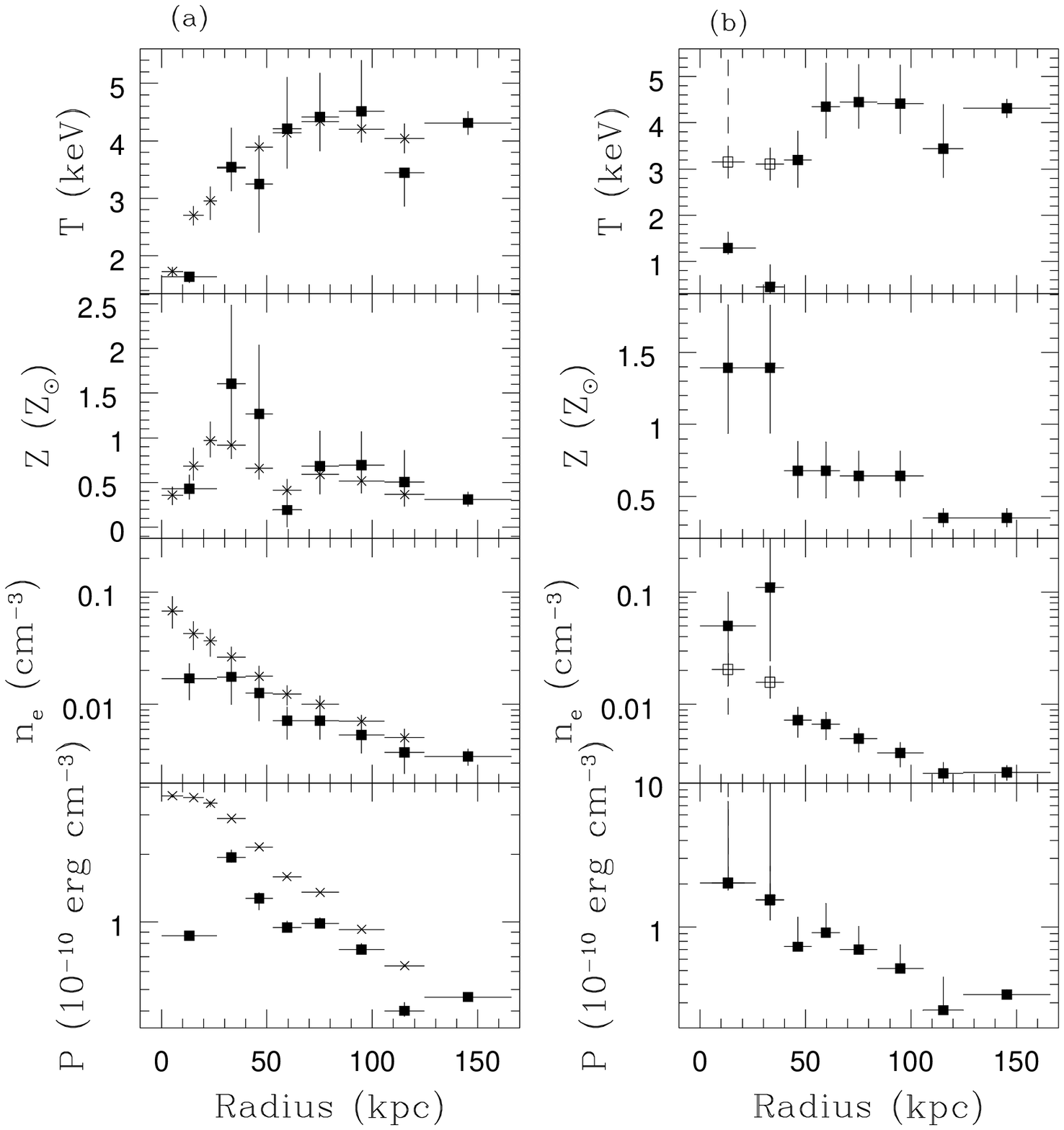,width=0.8\textwidth}
}
\caption{Deprojection analysis of the {\it Chandra}/ACIS-S data for A4059.  
  Panel (a) shows the results of fitting a single temperature plasma
  to the spectrum from each shell (filled squares), with the density
  determined from the emission measure assuming that the plasma
  uniformly fills the volume of the shell.  The crosses show the
  results from the (projected) annular study of
  Section~\ref{sec:proj}, with the density naively determined from the
  emission measure.  In panel (b), an additional plasma component has
  been included in the deprojected study in those bins for which it
  makes a significant improvement to the goodness of fit (i.e., the
  inner two bins). It is assumed that the two components are in
  pressure equilibrium, have the same metallicity, and jointly fill
  the volume of the shell.}
\label{fig:deproj}
\end{figure*}

To address this complication, we have performed a spectral analysis of
``deprojected'' spectra.  In detail, we deproject the cluster emission
into eight shells assuming spherical symmetry using the {\tt projct}
model within the {\sc xspec} spectral fitting package\footnote{Note
  that we use the latest version of {\tt projct} that can correctly
  handle the multiple datasets.}.  Clearly, any simple symmetry
assumption will break down in the morphologically complex inner
regions of A4059.  However, we might hope to perform a deprojection
analysis of this cluster beyond $30-40\kpc$, where it is fairly
regular.  

With this deprojection in hand, we initially model the spectrum of
each shell with an absorbed one-temperature {\tt mekal} model in which
the global abundance is a free parameter.  The density of the plasma
is determined from the plasma emission measure assuming that the
plasma uniformly fills the volume of the shell.  These results are
reported in Fig.~\ref{fig:deproj}a; for comparison, we also show the
results from fitting model-S to the spectra from the projected annuli
(naively computing the density from the observed emission measure of
the annulus).  It can be seen that the single-temperature deprojection
study reproduces the peak in metallicity at 30--40\,kpc.  Within this
radius, the spherical assumption clearly breaks downs and hence the
deprojection is not to be trusted.  Indeed, the leveling off of the
ICM density, and the drop in ICM pressure within the centralmost bin
is unphysical and almost certainly due to the morphological
complexities associated with the radio-galaxy/ICM interaction.

In order to examine the possibility of multiphase gas, we add an
additional temperature component to those deprojected spectra for
which it is a significant improvement in the goodness of fit
(employing the F-est with a 90\% level confidence threshold).  Only
the inner two radial shells required a second temperature component
(Fig.~\ref{fig:deproj}b).  As in the case of the projected study, the
central metallicity drop is removed by the addition of a second
component.  

\section{Discussion and conclusions}

\subsection{Summary of observational results}

There is clear evidence of a vigorous and complex
radio-galaxy/cluster interaction between PKS2354--35 and A4059.  Prior
to the analysis presented in this paper, the known facts relevant to
this interaction were:
\begin{enumerate}
\item There are two large ICM cavities approximately aligned with the
  axis of the radio galaxy.  Huang \& Sarazin (1998) and, later Heinz
  et al. (2002), showed that the radio source, as defined in the A-
  and B- array 4.8\,GHz and 8.5\,GHz VLA observations of Taylor et al.
  (1994) extends into the NW cavity, but does not extend to (or even
  point at) the SE cavity.
\item There is an offset between the center of the axis connecting the
  two cavities and the galactic nucleus.  One is given the impression
  that, assuming the cavities were created symmetrically by the radio
  galaxy, they have subsequently ``drifted'' in a NE direction.
\item There is a bright ridge of emission extending from the center of
  the cluster in the SW direction.  This ridge terminates about
  25\,kpc to the SW of the center in an abrupt edge.
\end{enumerate}
To this, we can now add the following informations:
\begin{enumerate}
\item VLA/CnB-array data taken at 1.4\,GHz, which is much better
matched to detecting arcmin-scale structures than the previous radio
data, still fails to detect any radio emission associated with the SE
X-ray cavity.
\item There is no indication that the gas around the X-ray cavities is
any hotter or higher entropy than the ambient gas.  In other words,
there is no evidence for a strong (or even moderately weak) shock
surrounding the X-ray cavities.
\item The SW ridge appears to be in approximate pressure balance with
the ambient material and is X-ray bright because of its lower
temperature and higher density.  The radiative cooling time in this
structure is much shorter than that of the surrounding ICM, becoming
as short as 100\,Myr (compared with a general ``core'' cooling time of
greater than 500\,Myr).
\item There is a robust metallicity gradient within the cluster, with
  high metallicity (approaching solar) in the cluster center and then
  declining by a factor of 2 beyond 50\,kpc.  This is reproduced in
  both the annular (i.e. projected) and deprojected spectral study.
  The presence of a central depression in the metallicity profile is
  suggested by single temperature fits to either the projected or
  deprojected spectra.  However, the reality of this feature is
  unclear (see above for details).
\item HST/WFPC-2 imaging reveals that the cD galaxy and host of
PKS2354--35, ESO349--G010, displays a prominent 5\,kpc dust lane
oriented roughly perpendicular to the radio-axis.  This suggests that
it has accreted a dust rich companion galaxy in the past $10^8\yr$ or
so.
\end{enumerate}
In this section, we discuss the constraints that these observations
place on the nature of the interaction.

\subsection{Inflating the cavities}

As discussed in Heinz et al. (2002), the current radio source is
likely too weak to produce notable cavities, and it is likely that the
observed ICM cavities are ``ghosts'' of a previous and more powerful
period of activity.  In this picture, the cavities are in a passive
phase of evolution (see Reynolds, Heinz \& Begelman 2002).  The X-ray
cavities, which were created by a past phase of supersonic lobe
expansion, have decelerated to sub-sonic velocities. Any shocks once
bounding the lobes have weakened into mere compression waves.  The
fact that this activity produces an expanding shell of ICM implies
that gas from the core regions will be lifted to higher points in the
cluster, thereby adiabatically cooling as it de-pressurizes.  This
cooling effect can largely offset the heating from the ICM
compression and (certainly to within the accuracy of our data) mask
any remaining signs of compressional heating.  This explains the lack
of hot gas in or around the cavities.

In this evolutionary phase, the cavities will buoyantly rise within
the cluster potential on a timescale a factor of a few longer than the
sound crossing time of $\sim 2 \times 10^7\,{\rm yrs}$. As they rise
buoyantly and expand, the relativistic electron population will
undergo synchrotron, inverse Compton, and adiabatic energy losses.
The synchrotron and inverse Compton losses result in a high-frequency
cut-off that gradually marches to lower and lower radio frequencies.
Using the standard formulae for synchrotron losses (e.g., Rybicki \&
Lightman 1979), it is readily shown that, assuming an isotropic
relativistic electron distribution evolving in a constant or
decreasing strength magnetic field, the high-frequency cut-off of the
synchrotron spectrum will obey
  \begin{equation}
    \nu_{\rm cut} \lesssim 26\,\left(\frac{B}{60\,\mu{\rm
    G}}\right)^{-3}\left(\frac{t}{20\,{\rm Myr}}\right)^{-2}\,{\rm MHz},
  \end{equation}
  where approximate equality corresponds to the case where the
  magnetic field and the particle pressure are constant in time.  This
  expression assumes no fresh injection or acceleration of
  relativistic electrons (which would turn the cut-off into a spectral
  break), and hence only applies once the radio-lobes are no longer
  supplied by active jets (i.e., after the radio-source ``dies''). The
  ICM pressure at the location of the ghost-cavities is measured to be
  approximately $p \approx 10^{-10}\,{\rm ergs\,cm^{-3}}$. If we
  assume that the synchrotron emitting plasma is in pressure
  equilibrium with the surrounding ICM (which is very likely to be
  true for the ghost cavities) and furthermore, that the magnetic
  field in the plasma has approximately equipartition strength and is
  tangled on scales small compared to the cavity size, this pressure
  gives us a field strength of $B \approx 60\,\mu{\rm G}$.  Thus,
  assuming ICM/cavity pressure balance and equipartition magnetic
  fields, we can see from eqn.(1) that the cavities will fade out of
  the 1.4 GHz band only 4 Myr or so after the outburst of the
  radio-galaxy activity has ceased. Since we believe the ghost
  cavities to be approximately 20 Myr old (Heinz et al.~2002), we see
  that there has been ample time for the plasma filling the cavities
  to fade out of the higher frequency radio bands {\it if the magnetic
    field posesses roughly equipartition strength}.
  
  Studies with ROSAT, {\em Chandra}, and {\em XMM-Newton} have allowed
  the magnetic field strengths of several radio lobes to be estimated
  through the direct detection of the X-rays thought to be produced by
  inverse Compoton scattering of the Cosmic Microwave Background (CMB)
  by the relativistic electrons (Leahy \& Gizani 2001; Hardcastle et
  al.  2002; Grandi et al. 2003; see also Wilson, Young, \& Shopbell
  2001 for related arguments in the hot spots of Cygnus A). In these
  studies, it is typically found that the magnetic field is at least a
  factor of two lower than the equipartition value.  Even if the
  magnetic field has half of the equipartition field strength, there
  is sufficient time for the 1.4\,GHz emission from the ghost cavities
  to fade.

Having put forward a fairly traditional hypothesis for the formation
and evolution of the X-ray cavities, we now proceed to consider the
complexities special to Abell~4059.

\subsection{Possible formation mechanisms for the SW ridge}

One of the most striking feature in the X-ray morphology of A~4059 is
the bright and cool SW ridge.  The SW edge of this ridge appears to be
surface across which the temperature and entropy of the gas change
significantly with little or no change in pressure.  In many ways,
this is similar to the ``cold fronts'' that have been observed in many
clusters (Markevitch et al. 2000; Vikhlinin, Markevitch \& Murray
2001).  Here, we will discuss four possible formation mechanisms for
this structure.

\subsubsection{A cool disk associated with a rotating cooling flow}

Huang \& Sarazin (1998), who were the first to note the SW ridge using
{\it ROSAT} HRI data, suggested that it might be the
rotationally-supported disk of cooled gas expected to form at the
center of a rapidly-rotating cooling flow.  The notion that such a
disk-like structure can form in high angular momentum cooling flows
has gained support from axisymmetric hydrodynamically simulations
(Garasi et al. 1998), although there are still unresolved questions as
to the effect that turbulent angular momentum transport may have on
the formation and stability of such disks (Nulsen, Stewart, \& Fabian
1984).

However, it is clear from the high-resolution {\it Chandra}-ACIS data
that the SW ridge does {\it not} extend NE of the cluster center,
i.e., it is one-sided.  This can be seen in both the total intensity
map (Fig.~\ref{fig:acis_image}) and, more clearly, in the temperature
map (Fig.~\ref{fig:maps}b).  This runs counter to the idea that the SW
ridge is part of a large ($\sim 20$\,kpc) disk at the center of the
cluster.  Thus, just on the basis of morphology, we can reject the
hypothesis that this structure is part of a disk associated with a
rotating cooling flow.

\subsubsection{Cool wakes of buoyantly rising radio plumes}

Numerical simulations of the buoyant phase of a radio-galaxies
evolution show that appreciable amounts of ICM from the cluster core
can become entrained in the ``wake'' of a buoyantly rising plume of
radio plasma (Br\"uggen et al. 2002; Reynolds, Heinz, \& Begelman
2002).  This material adiabatically decompresses and cools as it is
dragged upwards in the cluster potential, and would appear as distinct
filaments of cold and dense material strung out along the path of the
buoyant plume.  

As discussed by Young, Wilson \& Mundell (2002), these wakes of cold
gas are probably responsible for the arc-like feature seen in {\it
  ROSAT}-HRI and {\it Chandra}-ACIS observations of M87 and the core
of the Virgo cluster.  This structure is composed of narrow filaments
or columns of cold gas (with $kT\sim 1\keV$, compared with $kT\sim
3\keV$ for the surrounding ICM), probably in pressure equilibrium with
their surroundings, that extend for 2--3\,arcmins East and South-West
of M87.  They are coincident with, but more more narrowly confined
than, the 90\,cm radio arc observed by Owen, Eilek \& Kassim (2000).
This supports the idea that the filament has been entrained and pulled
out of the central parts of M87/Virgo by a buoyantly rising plume.

However, it seems unlikely that such a model can explain the SW ridge
of A~4059.  There is no indication of any radio-lobe (even a very old
one) in the SW direction, i.e., there is no radio emission and no ICM
cavity in that quadrant of the cluster.  Furthermore, the SW ridge
does not take on the form of a narrow filament reaching out from the
cD galaxy, as would be expected for wake material on the basis of both
the numerical simulations and the Young et al. (2002) observations of
Virgo.  Instead, the SW ridge is a rather broad and flaring feature
extending from the cD galaxy.  On the basis of these two observations,
we reject the hypothesis that the SW ridge corresponds to cool
material that has been entrained in the wake of a buoyantly rising
plume of radio-plasma.

\subsubsection{The accreted core of a cooler sub-cluster}

The discovery of a large dust lane in the HST-WFPC2 image of
ESO349-G010 suggests the accretion of a dust and gas rich companion
galaxy within the past $10^8$\,yrs.  It is possible that this
galactic-merger event was actually the late stages of the merger of a
smaller galaxy cluster or group with A~4059.  In this case, one may
attempt to identify the SW ridge with the remnant ICM core of the
minor cluster.  The well known correlation between the X-ray
luminosities and temperatures of galaxy clusters and groups, $L\sim
T^3$, means that the accreted minor system is likely to possess an ICM
that is significantly cooler than the ambient temperature of A~4059
($kT\sim 4\,keV$).  We note that similar ideas have been proposed to
explain the cold fronts observed in other clusters (Markevitch et
al. 2000; Bialek, Evrard \& Mohr 2002; Nagai \& Kratsov 2002).

However, this scenario may be problematic for the case of A~4059
(although it may well explain some of the classical cold fronts seen
in other clusters).  While it is true that the ICM {\it temperature}
of the accreted subcluster may initially be cooler than that of
A~4059, it will compress and heat as it enters the higher pressure
environment of the richer cluster.  The relevant thermodynamic
quantity to consider is the {\it entropy} of the ICM cores of A~4059
and the accreted cluster.  In fact, for clusters of the mass of A~4059
and smaller, the entropy of the ICM core is almost constant from one
cluster to another (Lloyd-Davies, Ponman \& Cannon 2000; Mushotzky et
al. 2003).  Thus, even if it evolved adiabatically, the ICM-core of
the accreted group would be compressionally heated to approximately
the ambient temperature of A~4059.  Any departure from adiabatic
evolution (e.g., the effects of shocks) will only increase the entropy
and temperature of the accreting ICM.  In order to produce a colder
region, radiative cooling needs to dominate the evolution of the
accreted core.  While this may be true, either fine tuning or
significant feedback (either via conduction or radio-galaxy heating)
is needed to prevent the core from cooling completely.

\subsubsection{A compression front associated with bulk ICM motion}

The apparent displacement of the center of the cavities from the
cluster center and the different position angle between the extended
radio emission and the cavities suggests bulk motion of the ICM flow.
Noting that the line connecting the centers of the two cavities misses
the radio-galaxy core by approximately 10\,arcsec, corresponding to
10\,kpc, we can use estimates of the age of the radio source ($\sim
20$\,Myr; Heinz et al. 2002) to estimate that the ICM is flowing past
the cD galaxy at a velocity of $500\kmps$ projected onto the plane of
the sky.  The different position angle between the axis connecting the
ghost cavities and the current extended radio emission demands either
a change in the radio-axis itself, or some rotation in the ICM flow.
Both numerical simulations (e.g., Roettiger, Loken \& Burns 1997) and
{\it Chandra} observations (e.g., Markevitch et al. 2003) suggest this
kind of large scale ICM ``sloshing'' can readily occur after a major
cluster merger.

In this picture, the bright and cool SW ridge is located at the
position where we expect the radio-galaxy induced expanding ICM shell
to be maximally compressed by the ICM flow.  The sharp SW edge of this
feature is readily interpreted at the interface between the ambient
ICM (which we suppose is flowing in a NE direction) and the expanding
ICM shell formed by the same period of radio-galaxy activity that
formed the X-ray cavities.

One might expect that such compression would heat this material,
contrary to observations.  However, the fact that the cooling time of
the ridge material is small demands that we consider radiative cooling
effects.  In the simplest case of adiabatic compression in the
bremsstrahlung regime, the cooling time is proportional to $n^{-2/3}$.
Hence, a weak shock will slightly reduce the cooling timescale.
Radiative cooling can be further aided by the kinematics of the
radio-galaxy/cluster interaction, which keeps this material in the
high pressure regions of the cluster core for longer.  Even given
this, there appears to be a fine tuning problem; it is difficult to
explain the cooling of the SW ridge unless it was on the verge of
undergoing dramatic radiative cooling anyways.  A detailed exploration
of these hydrodynamical and radiative questions will be deferred until
future publications.

\section*{acknowledgments}

We thank Mitchell Begelman, and Andy Young for stimulating
conversations throughout the course of this work.  We are also
grateful to the anonymous referee for comments that significantly
improved the paper.  This work was supported by SAO grant GO0-1129X
(YC), the National Science Foundation grant AST0205990 (CSR), and
Korea Research Foundation grant KRF 2001-041-D00052(JY). Additional
support for this work (SH) was provided by the National Aeronautics
and Space Administration through Chandra Postdoctoral Fellowship Award
Number PF3-40026 issued by the Chandra X-ray Observatory Center, which
is operated by the Smithsonian Astrophysical Observatory for and on
behalf of the National Aeronautics Space Administration under contract
NAS8-39073.

\end{document}